\documentclass[a4paper]{jpconf}
\usepackage{amssymb,amsmath}
\usepackage[dvips]{graphicx}
\usepackage{soul}
\newcommand{\vp}{\varphi}

\begin{document}

\title{Influence of Josephson current second harmonic on stability of magnetic
flux in long junctions}

\author{P.Kh.Atanasova$^{1}$, \frame{T.L.Boyadjiev},\\
 Yu.M.Shukrinov$^{1}$, E.V.Zemlyanaya$^{1}$, P. Seidel$^{2}$}

\address{$^1$ Joint Institute for Nuclear Research, Dubna 14198, Russia.}
\address{$^2$  Institut f\"{u}r Festk\"{o}rperphysik, Friedrich-Schiller-Universit\"{a}t Jena, D-07743 Jena, Germany}
\ead{poli@jinr.ru}

\begin{abstract}
{\it}We study the long Josephson junction (LJJ) model which takes into account the second harmonic of the
Fourier expansion of Josephson current. The dependence of the static magnetic flux distributions on parameters
of the model are investigated numerically. Stability of the static solutions is  checked by the sign of the smallest
eigenvalue of the associated  Sturm-Liouville problem. New solutions which do not exist in the traditional
model, have been found. Investigation of the influence of second harmonic on the stability of magnetic flux
distributions for main solutions is performed.
\end{abstract}

\section{Motivation and model}

The physical properties of the magnetic flux in Josephson junctions (JJ) are the contemporary superconducting electronics base. For
the traditional JJ the current-phase dependence  is a sinusoidal function. Such a model is described by the
sine-Gordon equation. For a sufficiently wide class of JJ the superconducting Josephson current as a function of
magnetic flux $\vp$  can be represented as a sine series \cite{gki04,l85}:
\begin{equation}\label{i_s_full}
    I_S = I_c \sin \varphi + \sum_{m=2}^{\infty} I_m \sin m
    \varphi\,.
\end{equation}

It was recognized recently that the higher harmonics in this expansion are important in many applications, in
particular, in junctions like SNINS and SFIFS, where S is a superconductor, I is an insulator, N is a normal metal and F is a weak metallic  ferromagnet
\cite{gki04,ror01}. The interesting properties of LJJ with an arbitrarily strong amplitude
of second harmonic in current phase relation were considered in \cite{goldobin07}.

Using only first two terms of the expansion (\ref{i_s_full}) leads to  the double {sine}-Gordon
equation (2SG) \cite{l85}.

\begin{equation}\label{2sg}
    -\varphi\,'' + a_1 \sin \varphi + a_2 \sin 2\varphi -\gamma = 0\,, x \in (-l;l)\,.
\end{equation}
Here and below the prime means a derivative with respect to the coordinate $x$. The magnitude $\gamma$ is the
external current, $l$ is the semilength of the junction,  $a_1$ and $a_2$ are the normalized amplitudes of the
first and second harmonics of the Josephson current \cite{gki04,bk03}.  All the magnitudes are dimensionless.

The boundary conditions for (\ref{2sg}) have the form
\begin{equation}\label{bound2sg}
    \varphi\,'(\pm l) = h_e,
\end{equation}
where $h_e$ is  external magnetic field.


Numerical solution of the nonlinear boundary problem (\ref{2sg}), (\ref{bound2sg}) is solved on
the basis of the continuous analog of Newton's method [7].

Stability and bifurcations of static solutions $\vp(x,p)$, where $p = (l,a_1,a_2,h_e,\gamma)$ are analyzed on the basis of numerical solution of
the corresponding Sturm-Liouville problem \cite{pbvzpc07}:
\begin{equation}
\label{slp}
   -\psi\,'' + q(x)\psi = \lambda \psi, \quad
    \psi\,'(\pm l) = 0, \quad q(x) = a_1 \cos \varphi + 2 a_2 \cos 2 \varphi.
\end{equation}
The minimal eigenvalue  $\lambda_0(p) > 0$ corresponds to the stable solution. In case $\lambda_0(p) < 0$ solution $\vp(x,p)$ is unstable. The case $\lambda_0(p) = 0$ indicates the bifurcation with respect to one of the parameters $p$. We characterize the solutions of equation
(\ref{2sg}) by number of fluxons $N(p)$ which is defined as
\begin{equation}\label{nof}
 N(p) = \frac{1}{2l \pi} \int\limits_{-l}^l \varphi(x)\,dx\,.
\end{equation}

\section{Deformation of the Meissner solution $M_0$}

\begin{figure}[ht]
\includegraphics[width=17pc]{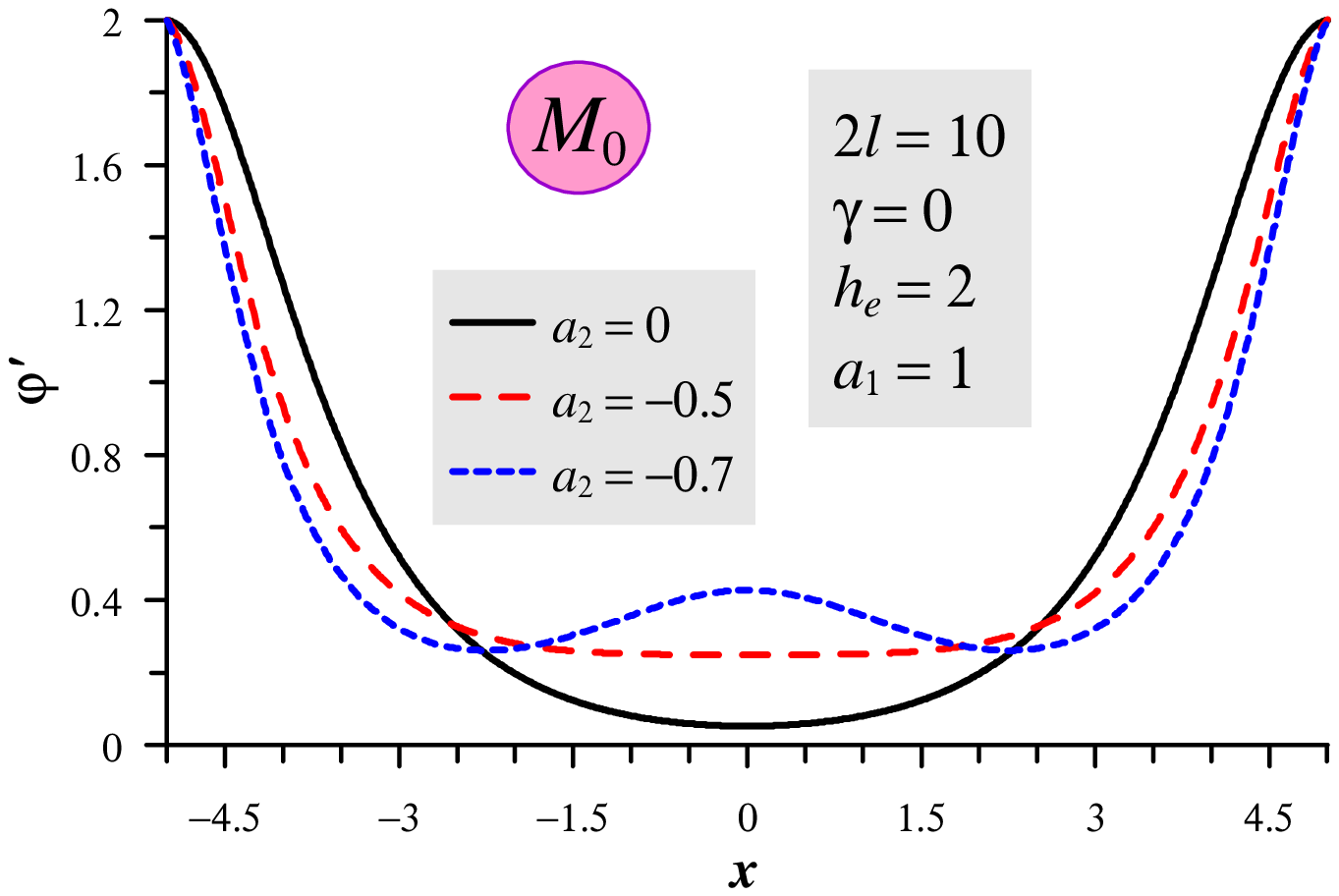}
\includegraphics[width=17pc]{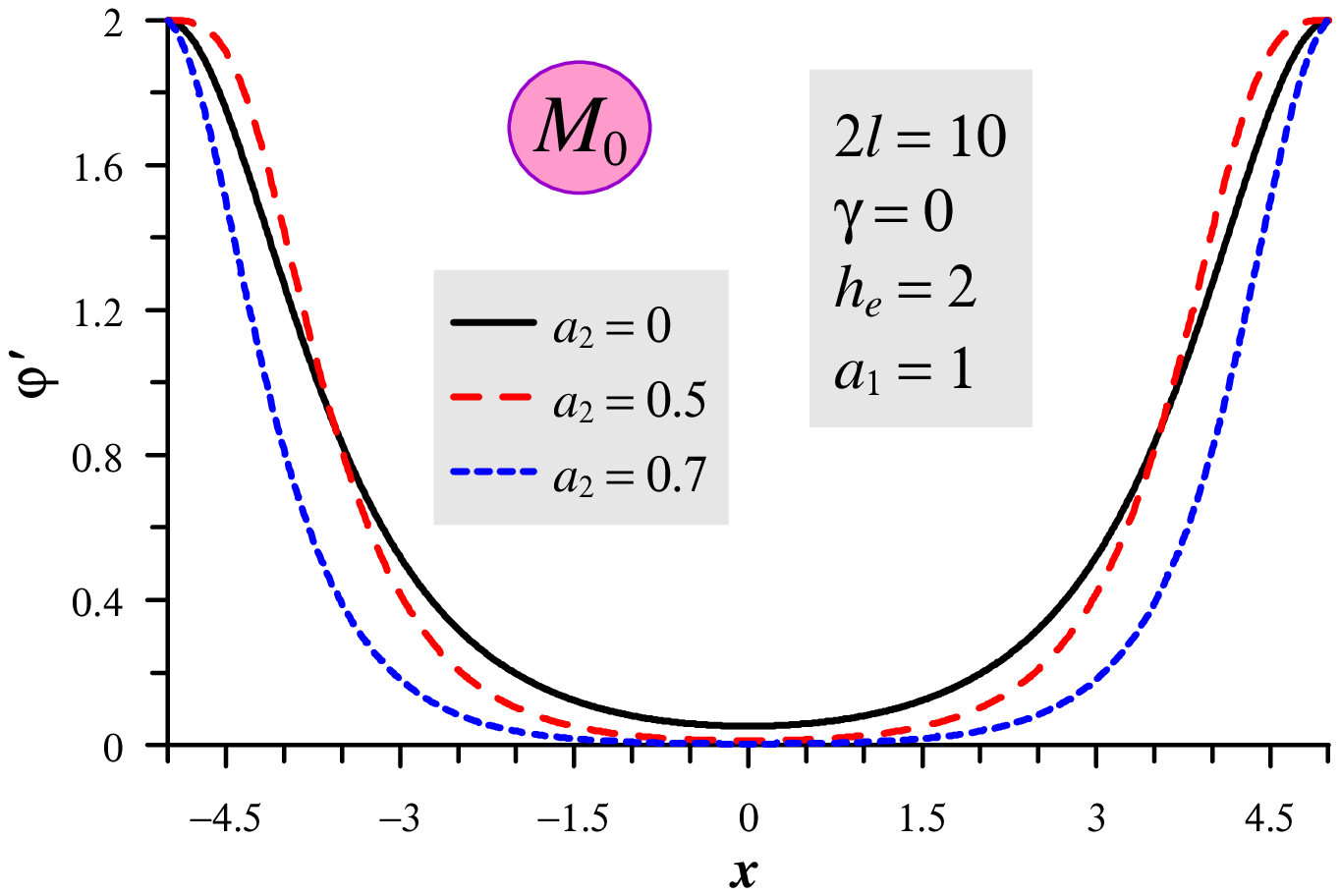}
\caption{The internal magnetic field of the Meissner solution $M_0$  for LJJ with $2l = 10$, $\gamma = 0$, $h_e
= 2$ and $a_1 = 1$ at different values of the parameter $a_2$. Left figure shows $\varphi\,'$ at negative $a_2$,
right one shows at positive $a_2$. }\label{Sols_M0}
\end{figure}

In the ``traditional'' case $a_2 = 0$ two trivial solutions $\varphi = 0$  and $\varphi = \pi$  of (\ref{2sg}),
(\ref{bound2sg})  are known  at $\gamma = 0$ and $h_e = 0$, which are denoted by $M_{0}$ ($N[M_0] = 0$) and
$M_{\pi}$ ($N[M_{\pi}] = 1$), respectively.  Accounting of the second harmonic $a_2 \sin 2\varphi$ leads to the appearing of two
additional solutions $\varphi = \pm \arccos (-a_1/2a_2)$ denoted as $M_{\pm ac}$ ($N[M_{\pm ac}]$ are not integer numbers and
depend on the value of second harmonic). Stability properties of trivial
solutions in dependence on parameter $a_2$ are considered in \cite{azbsh10} and \cite{borovetz}.

All solutions with $N[\vp] = 0$  we denote here by $M_{0}$, even they
are changed by the influence of the external magnetic field and parameter $a_2$.Within the LJJ model, the basic Meissner solution  $M_{0}$ demonstrates the screening of the external magnetic
field. Deformation of $M_{0}$ at different values of parameter $a_2$ under the external magnetic field $h_e=2$
is shown in Fig.\ \ref{Sols_M0}.  At $a_2<0$ (left panel) the screening effect diminishes when the coefficient
$a_2$ decreases and it amplifies at $a_2>0$ (right panel) with growing of $a_2$.

\section{Deformation of the fluxon solutions $\Phi^n$}

\begin{figure}[ht]
\includegraphics[width=17pc]{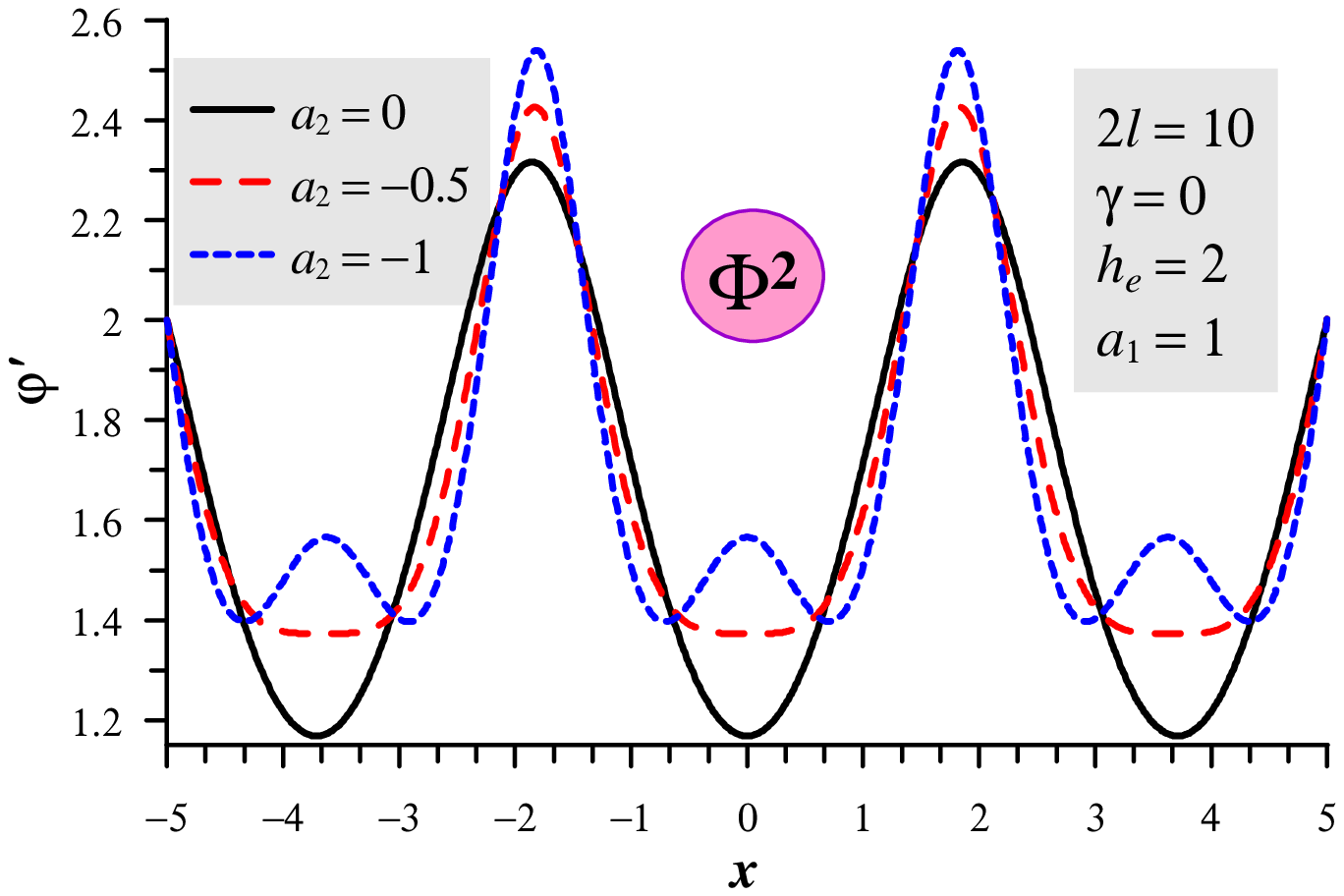}
\includegraphics[width=17pc]{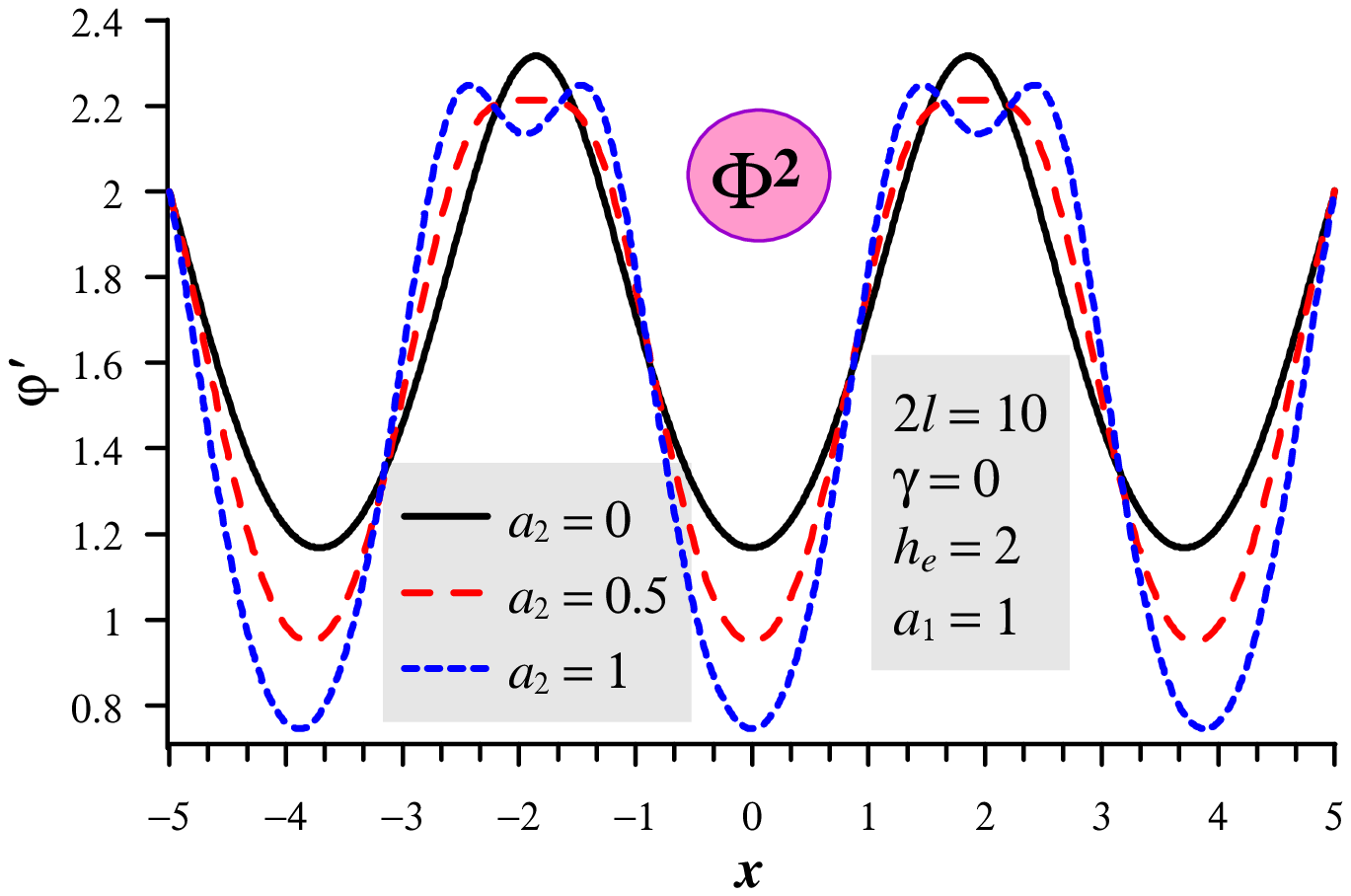}
\caption{The internal magnetic field of the distribution $\Phi^2$  at $2l = 10$, $\gamma = 0$, $h_e = 2$ and
$a_1 = 1$ at different values of the parameter $a_2$. Left figure shows $\varphi\,'$ at negative $a_2$, right one
shows at positive $a_2$.}\label{Sols_F2}
\end{figure}

The solutions with $N[\vp] = 1$ which are not   $\vp =\pi$ at $h_e = 0$ and
$\gamma = 0$ we denote as $\Phi^1$. The deformation of the $\vp\,'(x)$ of fluxon states under influence of the parameter $a_2 \in [-1;1]$ was
considered in \cite{borovetz}. We observe a qualitative change in the local minima with decrease of $a_2$ in
the interval [-1;0] and in the local maxima with increase of $a_2$ in the interval [0;1].  With change of the
coefficient $a_2$ the number of fluxons
corresponding to the distribution $\Phi^1$ is conserved \cite{pbvzpc07} i.e. $ {\partial N}/{\partial a_2 } = 0$
and $N[\Phi^1] = 1$.  As external magnetic field $h_e$ is growing, more complicated stable fluxon
with $n = N[\vp] = 2,3,\ldots$  appear
which
are denoted by $\Phi^n$.

The effect of the second harmonic contribution on $\Phi^1$ was studied in \cite{goldobin07} and \cite{lozenetz}. It was shown in \cite{goldobin07} that taking into account at $h_e = 0$ the second harmonics in current
phase relation with $a_2 < -0.5$ leads to the appearance of the ``small'' fluxon state additionally to the
traditional ``large'' one. Taking into account  \eqref{nof}, we call
the solution at $a_2 = -0.7$ in Fig.~\ref{Sols_M0}, left panel as  $M_0$ (in \cite{goldobin07} it is a
``small''  fluxon), because $N[M_0] = 0$. Number of fluxons of ``large'' fluxon is equal to $N[large] = 1$ and  we
denote it by $\Phi^1$.  In \cite{lozenetz}, stability properties of ``large'' fluxon $\Phi^1$ at nonzero $h_e$ have been studied. Two coexisting stable $\Phi^1$  like fluxons were demonstrated in some
region of magnetic field. Here, we
investigate this effect in case of two-fluxon and three-fluxon distributions. The relation between the ``small'', ``large'' fluxons and trivial
solutions under the influence of the external magnetic field and the
second harmonic are the point of our further research.

In Fig.~\ref{Sols_F2} the internal magnetic field of the two-fluxon distribution $\Phi^2$  for $2l = 10$, $\gamma = 0$,
$h_e = 2$ and $a_1 = 1$ at different values of the parameter $a_2$ is presented. We observe a strong deformation
of the fluxon distributions with decrease of the parameter $a_2$ in the interval [-1;0]. At $a_2 = -0.5$ the
curve of internal magnetic field $\vp\,'(x)$ has a plateau at the points $x \approx -4$, $x \approx 0$ and $x \approx 4$ (Fig.\
\ref{Sols_F2}, left panel). Further increase of the absolute value of $a_2$ leads to the transformation of these
plateaus to local maxima  of the internal magnetic field. Thus, accounting of the $a_2$ contribution qualitatively changes a
shape of the fluxon distribution $\Phi^2$. A similar  deformation in the local maxima regions (points $x \approx
-2$ and $x \approx 2$) is observed for  $a_2 > 0$ (Fig.\ \ref{Sols_F2}, right panel). We stress that the number
of fluxons is conserved $N[\Phi^2] = 2$ for all values $a_2$.

\section{Stability analysis of the static fluxon distribution {$\Phi^n$}}

\begin{figure}[ht]
\includegraphics[width=17pc]{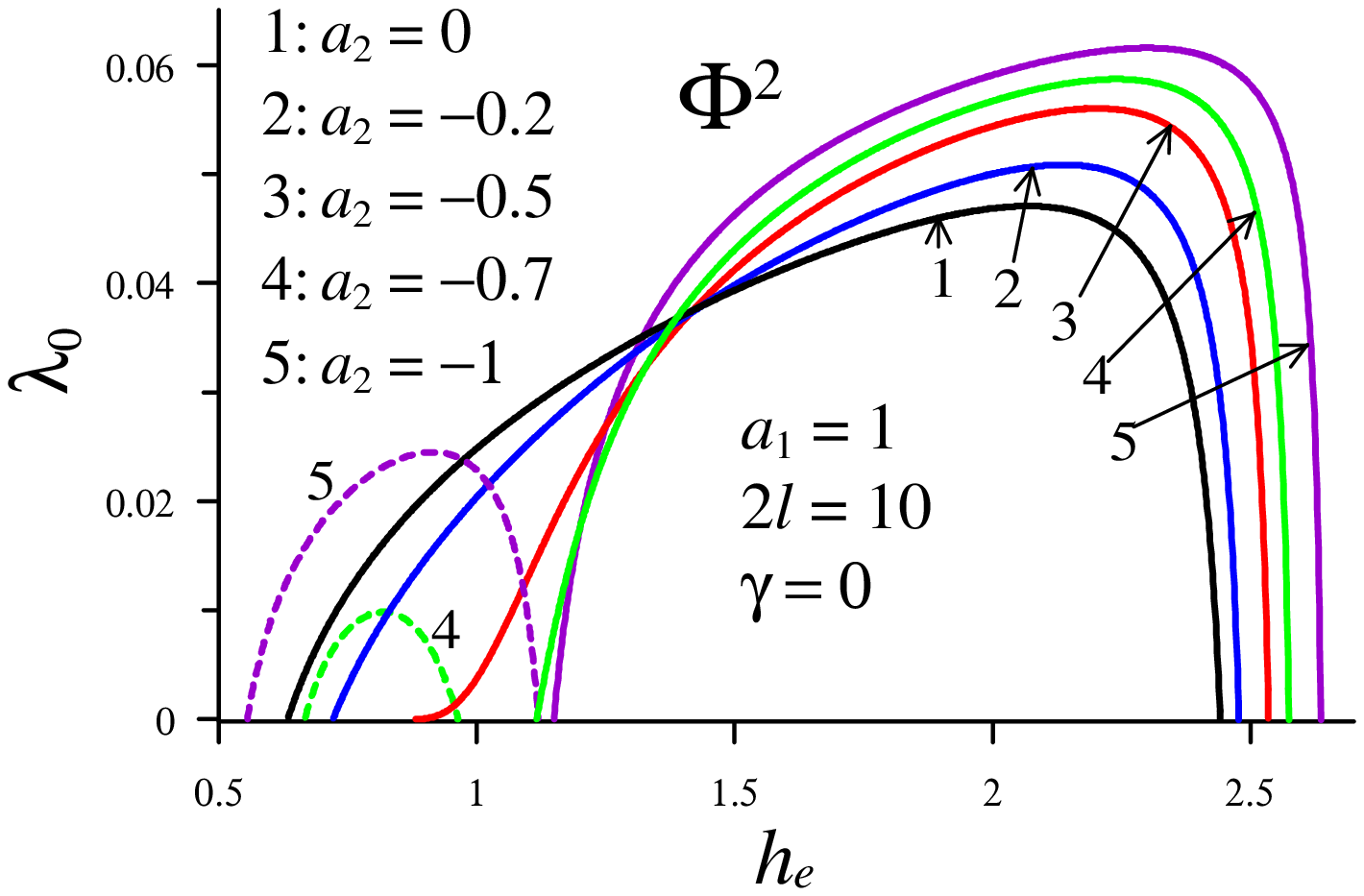}
\includegraphics[width=17pc]{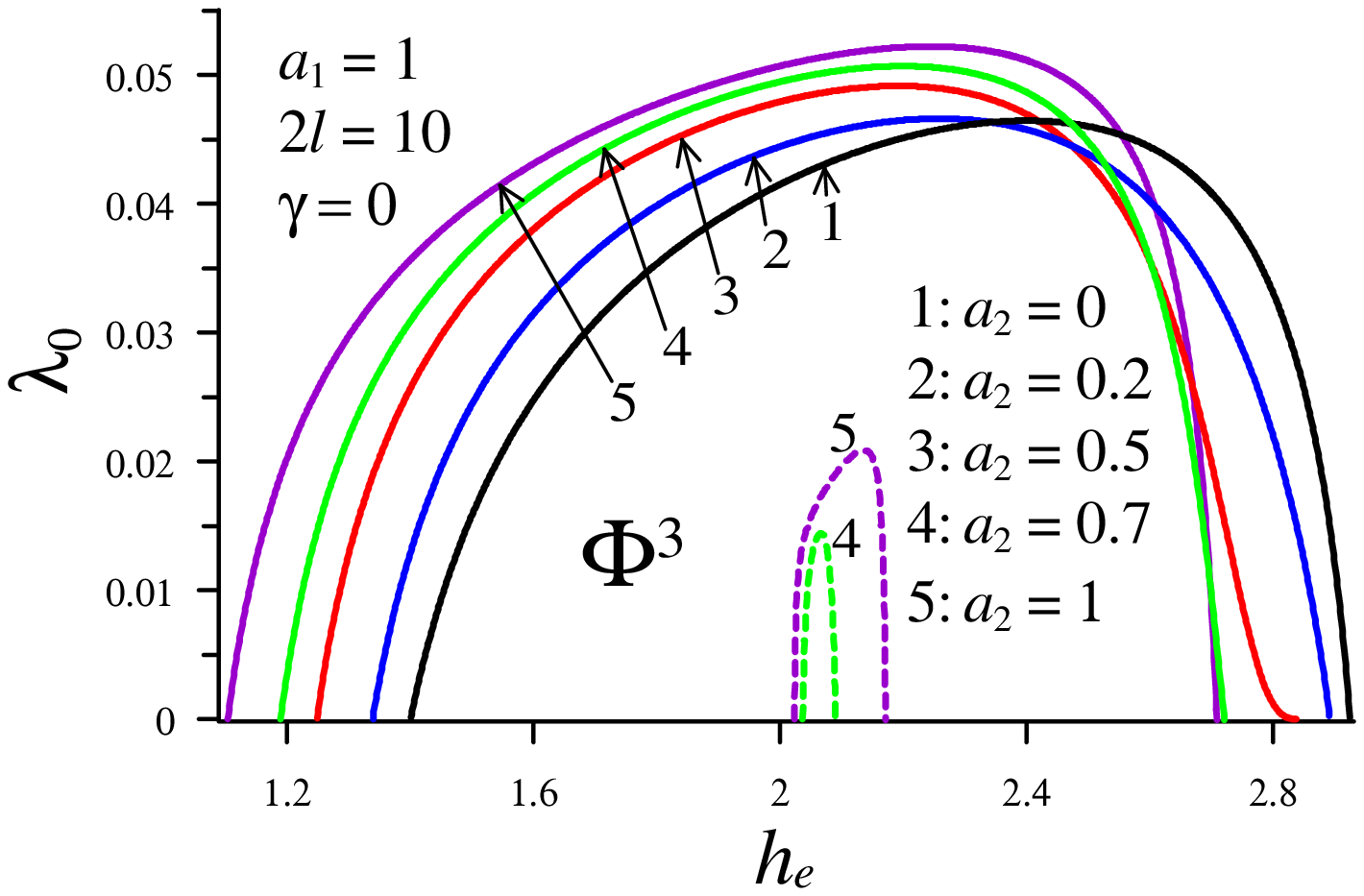}
\caption{The dependence $\lambda_0(h_e)$  for $\Phi^n$  at $2l = 10$, $\gamma = 0$, $h_e = 2$ and $a_1  = 1$ at
different values of the parameter $a_2$. Left figure shows $\lambda_0(h_e)$  for $\Phi^2$ at negative $a_2$, right one shows $\lambda_0(h_e)$  for $\Phi^3$ at
positive $a_2$.}\label{EV0_he_F2}
\end{figure}

The two-fluxon solution {$\Phi^2$} in LJJ with $2l=10$ and $\gamma=0$ is stable at $0.5 \lesssim h_e  \lesssim
2.5$. Qualitatively, the behavior of the curves  $\lambda_0(h_e)$ for this solution is similar to the case of
one-fluxon solution $\Phi^1$ \cite{lozenetz}. The only difference is that in the {$\Phi^2$} case we don't have
a region of coexistence of two stable branches that was observed for negative $a_2$ in the {$\Phi^1$} case.

Results for $a_2 \in [-1;0]$ are presented in Fig.~\ref{EV0_he_F2}, left panel. When  $a_2$  decreases in
$(-0.5;0]$, the curve $\lambda_0(h_e)$  moves to the right. At  $a_2 < -0.5$  the curve $\lambda_0(h_e)$
corresponding to the stable solution $\Phi^2$ has two separate branches.

The $\lambda_0(h_e)$ curves for $\Phi^3$ distribution for $a_2 \in [0;1]$  are demonstrated in
Fig.~\ref{EV0_he_F2}, right panel. It is seen, when $a_2$ is growing in the interval $[0;1]$, the first
bifurcation point moves to the left. Contrary, at $a_2>0.7$ the bifurcation point moves to the right. The second
bifurcation point moves to the left as $a_2$ is growing from 0 to $a_2 \approx 0.7$. Contrary, at $a_2>0.7$ the second bifurcation point moves to the right. For three-fluxon state the following effect
is observed. At $a_2
> 0.5$ new curves appear and we observe an existence of two different 3-fluxon states simultaneously.

For trivial solutions our calculations show the reducing of Meissner screening when the second harmonic is negative. New fluxon solutions which appear in case $a_2 \neq 0$ have a probability to be observed in the experiment. So, it would be interesting to test this.

As summary we note that our numerical investigations show that accounting of the second harmonic contributions
significantly changes the shape and stability properties of trivial and fluxon static distributions in LJJ.

We  thank to E. Goldobin for the stimulating discussions and important suggestions. This research was
partially supported by Heisenberg-Landau Program.
 The work of  P.Kh.A. is partially supported in the frame of the Program for collaboration of JINR-Dubna and Bulgarian scientific centers
``JINR -- Bulgaria''. E.V.Z. was partially supported by  RFFI under grant 09-01-00770-a.  P.Kh.A. and E.V.Z. are
thankful to I.V.Puzynin and T.P. Puzynina for support of this research.


\end{document}